\documentclass[aps,physrev,twocolumn,groupedaddress,10pt]{revtex4-2}

\usepackage{graphicx}
\usepackage{multirow}
\usepackage{float}
\usepackage{hyperref}
\begin{document}


\title{Scoring-Assisted Generative Exploration for Proteins (SAGE-Prot): A Framework for Multi-Objective Protein Optimization via Iterative Sequence Generation and Evaluation}



\author{Hocheol Lim}
\email{ihc0213@yonsei.ac.kr}  
\affiliation{Bioinformatics and Molecular Design Research Center (BMDRC), Incheon, Republic of Korea}

\author{Geon-Ho Lee}
\affiliation{Bioinformatics and Molecular Design Research Center (BMDRC), Incheon, Republic of Korea}

\author{Kyoung Tai No}
\email{ktno@yonsei.ac.kr}
\affiliation{Bioinformatics and Molecular Design Research Center (BMDRC), Incheon, Republic of Korea}


\date{\today}

\begin{abstract}
Proteins play essential roles in nature, from catalyzing biochemical reactions to binding specific targets. Advances in protein engineering have the potential to revolutionize biotechnology and healthcare by designing proteins with tailored properties. Machine learning and generative models have transformed protein design by enabling the exploration of vast sequence-function landscapes. Here, we introduce Scoring-Assisted Generative Exploration for Proteins (SAGE-Prot), a framework that iteratively combines autoregressive protein generation with quantitative structure-property relationship models for fine-tuned optimization. By integrating diverse protein descriptors, SAGE-Prot enhances key properties, including binding affinity, thermal stability, enzymatic activity, and solubility. We demonstrate its effectiveness by optimizing GB1 for binding affinity and thermal stability and TEM-1 for enzymatic activity and solubility. Leveraging curriculum learning, SAGE-Prot adapts rapidly to increasingly complex design objectives, building on past successes. Experimental validation demonstrated that SAGE-Prot–generated proteins substantially outperformed their wild-type counterparts, achieving up to a 17-fold increase in $\beta$-lactamase activity, underscoring SAGE-Prot’s potential to tackle critical challenges in protein engineering. As generative models continue to evolve, approaches like SAGE-Prot will be indispensable for advancing rational protein design.
\end{abstract}


\maketitle

\section{\label{sec:intro} Introduction}
Proteins perform diverse roles, including catalyzing biochemical reactions, providing structural support, and regulating biological processes. Their functions depend on amino acid sequences, which define three-dimensional structures and influence properties like binding affinity, stability, solubility, and catalytic efficiency. The sequence space for proteins is vast; a 100-amino-acid protein has approximately 10\textsuperscript{130} possible combinations, but only a small subset is functional. Protein design and engineering tackle this challenge by creating or modifying proteins with desired properties for applications in medicine, biotechnology, agriculture, and environmental sustainability. These advancements have enhanced natural proteins and enabled the creation of proteins with novel functions, driving innovation across various fields.

Protein design and engineering rely on two primary strategies: directed evolution and rational design. Directed evolution introduces random mutations and selects improved variants through iterative screening, yielding successes in enzyme design and therapeutic antibody development \cite{ref01}. However, this method is labor-intensive and limited by the randomness of mutations. Rational design uses structural and functional insights to introduce targeted sequence changes, supported by computational tools like molecular dynamics and protein folding models \cite{ref02}. Despite its precision, this approach is constrained by limited structural data and the complexity of protein dynamics. Semi-rational design bridges these gaps by combining the precision of rational design with the exploratory capacity of directed evolution \cite{ref03, ref04}. By targeting mutations to key structural or functional regions, it optimizes protein properties efficiently. This hybrid strategy has been particularly effective in enhancing functional diversity and protein performance.

Machine learning (ML) has revolutionized protein design and engineering by enabling efficient exploration of the vast sequence-function landscape. ML models analyze large datasets of protein sequences, structures, and properties to predict how mutations affect attributes such as binding affinity, stability, solubility, and catalytic efficiency \cite{ref05}. These models use a range of descriptors, from simple mutation indicators to advanced sequence embeddings, to capture both local and global mutation effects \cite{ref06, ref07, ref08, ref09, ref10, ref11, ref12}. One prominent ML application is optimizing sequence-function landscapes. Supervised learning models, such as regressors, predict variants with enhanced properties, validated in experiments like improving antibody binding in protein G domains beyond training data \cite{ref13}. Deep generative models, including variational autoencoders, autoregressive, and diffusion models, further advance the field by generating novel protein sequences while preserving functionality and structural integrity \cite{ref14, ref15, ref16, ref17}. These models leverage biochemical and evolutionary constraints to explore uncharted sequence space. ML-guided approaches significantly reduce reliance on labor-intensive trial-and-error methods, enabling the creation of proteins with novel functions, improved stability, and greater therapeutic potential.

In this study, we developed Scoring-Assisted Generative Exploration for Proteins (SAGE-Prot), a systematic framework for optimizing protein properties through iterative fine-tuning, combining protein sequence generation and evaluation to create novel proteins for targeted applications. SAGE-Prot employs natural language processing (NLP) models, such as Long Short-Term Memory (LSTM) and Transformer Decoder (TD), pre-trained on curated protein sequence datasets to generate diverse sequences via autoregressive modeling. To further enhance sequence diversity, genetic algorithms (GA) introduce variations through operations such as insertion, deletion, and substitution mutations, sequence merging via alignment, and retrieval-augmented generation using homolog search. Generated sequences are evaluated with scoring models based on comprehensive protein descriptors, predicting their properties using quantitative structure-property relationship (QSPR) models. Using SAGE-Prot, we improved the binding affinity and thermal stability of Protein G domain $\beta$1 (GB1) from \textit{Streptococcus group G} and enhanced the enzymatic activity and solubility of TEM-1 $\beta$-lactamase (TEM-1) from \textit{Escherichia coli}. This iterative refinement process enables SAGE-Prot to achieve desired protein characteristics effectively. Experimental validation of the top-ranked TEM-1 variants confirmed enhanced properties compared to wild-type, highlighting SAGE-Prot's potential as a robust tool for engineering proteins with tailored properties for diverse real-world applications in protein engineering.

\section{\label{sec:methods} Methods}

\subsection{Scoring-Assisted Generative Exploration for Proteins (SAGE-Prot)}
Scoring-assisted generative exploration (SAGE) employs an iterative process that alternatives between molecule generation and evaluation \cite{ref18, ref19}. The generation step is performed using autoregressive NLP models and diversification operators, while the evaluation step leverages various scoring models to align the generated molecules with specific desired properties. The SAGE framework, which has been applied to optimize the properties of chemicals \cite{ref18, ref19} and ionic liquids \cite{ref20}, was extended to proteins (SAGE-Prot) by pre-training Long Short-Term Memory (LSTM) \cite{ref21, ref22} and Transformer Decoder (TD) \cite{ref23} models and integrating protein-specific QSPR scoring models.

Proteins are represented as sequences of amino acids, using a total of 31 tokens. These tokens include 20 for canonical amino acids, as well as tokens for representing the sequence start and end, padding, ambiguous amino acids (B for Asx, Z for Glx, and X for Any), and sequence gaps for insertions and deletions. The protein sequences, used for pre-training the SAGE-Prot models, were obtained from SWISS-PROT \cite{ref24} and NCBI-BLAST \cite{ref25}, as summarized in Table~\ref{tab:table_1}. To avoid similarity with target protein drugs in the benchmark, the protein sequences with a maximum similarity greater than 0.5 were excluded, as determined by pairwise sequence alignment with the BLOSUM62 substitution matrix in Biopython \cite{ref26}. This process resulted in reduced protein datasets (SwissProt-reduced). Custom datasets for TEM-1 $\beta$-lactamase were constructed using NCBI-BLAST (ver. 2.15.0+) by querying the non-redundant dataset (ver. 2024.01.17) with \textit{Escherichia coli}-derived TEM-1 $\beta$-lactamase (UniProt ID: P62593) sequences at an e-value threshold of 10. These three datasets were then randomly divided into training and validation sets in proportions of 0.882 and 0.118, respectively.

The autoregressive NLP components in SAGE-Prot for pre-training include LSTM and TD models. The LSTM model features three layers with 1024 hidden units, a dropout rate of 0.2, a learning rate of 0.001, and a batch size of 256. The TD model incorporates four attention heads, three decoder layers, 512 hidden units, a dropout rate of 0.2, an embedding size of 128, a learning rate of 0.001, and a batch size of 128. The LSTM and TD models were pre-trained using the Adam optimizer \cite{ref27}, each for 150 epochs across three datasets. The best model weights were selected based on minimum validation loss.

The proteins generated by the pre-trained NLP models were evaluated using several metrics: validity, length, uniqueness, and novelty. Validity refers to the proportion of generated protein sequences that consist exclusively of the 20 canonical amino acids. This metric assesses the model's ability to generate accurate protein sequences without relying on special tokens for ambiguous amino acids or sequence gaps. Length represents the average length and standard deviation of the protein sequences containing only canonical amino acids, providing insight into the typical lengths of proteins produced by the model. Uniqueness evaluates the model's ability to generate a diverse set of protein sequences, avoiding repetitive or limited outputs. Novelty is assessed by calculating the proportion of generated protein sequences that are not present in the training dataset.

The pre-trained NLP models generate protein sequences in each iteration, and the generated proteins are first verified to ensure they consist solely of canonical amino acids. Any special tokens other than canonical amino acids are either excluded or replaced with canonical amino acids. If tokens representing ambiguous amino acids are present, they are randomly replaced with one of the canonical amino acids they represent, with equal probability. Subsequently, protein variation operators such as homolog search, mutation, and crossover are applied with probabilities of 1\%, 1\%, and 98\%, respectively. Each operator is iterated up to 10 times to ensure that the resulting sequences differ from the query sequence. In the homolog search step, the generated protein is compared against the landmark dataset (ver. 2024.01.17.) using NCBI-pBLAST (ver. 2.15.0+), and one of the top 10 ranking homologous sequences is selected randomly. The mutation operator introduces insertion, deletion, and substitution (non-synonymous) mutations at the amino acid level. A total of 14 mutation operators are used, each selected randomly with equal probability. These consist of one insertion, one deletion, and twelve substitutions. The twelve substitution operators involve replacing amino acids within predefined groups: positive, negative, aromatic, aliphatic, polar, nonpolar, DN-pair, EQ-pair, small, charged, neutral, and all amino acids. Substitutions occur within these groups, ensuring the replacement amino acid belongs to the same group as the original. The crossover operator performs sequence alignment of two protein sequences using Biopython (ver. 1.81) \cite{ref26} based on the BLOSUM62 matrix. From the aligned sequences, a token is randomly selected at each position, including sequence gaps. The resulting sequence, with gaps removed, combines the two protein sequences while preserving the alignment regions to create a new protein sequence.

After protein variation, the proteins are once again checked to ensure whether they comprise canonical amino acids. The proteins are then ranked based on their scores, and a fixed number of top-ranked proteins are selected for fine-tuning the NLP via a storage buffer at each step. Similar to SAGE \cite{ref18} and SAGE-IL \cite{ref20}, the top-ranked 1024 proteins in the storage buffer are preserved throughout the entire process. These top-ranked proteins are utilized for fine-tuning, with optimization performed using the Adam optimizer at a learning rate of 0.001 and a batch size of 256 over 8 epochs. NLP-based models (LSTM and TD) generate 16,384 proteins. In contrast, the GA-only model randomly selects 16,384 proteins from the training data at the start and generates 16,384 proteins. Hybrid NLP/GA models (LSTM/GA and TD/GA) produce 8192 proteins. For benchmarking, 100 iterations of SAGE-Prot were conducted across rediscovery, similarity, SPO, and MPO tasks.

\subsection{Goal-Directed Benchmarks and Score Definition in SAGE-Prot)}

Goal-directed benchmarks for evaluating generation performance followed protocols from the SAGE \cite{ref18} and SAGE-IL \cite{ref20} studies. In rediscovery tasks, the goal was to identify specific target proteins, including insulin, parathyroid hormone, interferon-$\gamma$, interferon-$\beta$, interferon-$\alpha$2, erythropoietin, caplacizumab, pexelizumab, asparaginase, and thrombopoietin. In similarity tasks, the objective was to generate proteins closely resembling these targets. Both tasks utilized sequence alignment with BLOSUM62 to measure identity and similarity scores between target proteins and generated proteins. Identity was assessed by checking whether aligned positions had the same amino acid, while similarity was determined by whether aligned amino acids had a positive BLOSUM62 substitution score. These values were summed up and normalized by the length of the aligned sequence. Additionally, we calculated the coverage ratio, representing the proportion of the aligned region relative to the aligned sequence length of the query and target. The rediscovery score was obtained by multiplying the identity value by the coverage ratio, while the similarity score was derived by multiplying the similarity value by the coverage ratio.

Single-property optimization (SPO) tasks aim to enhance protein properties such as binding affinity, thermal stability, enzymatic activity, and protein solubility in proteins. These tasks also consider scores based on protein length and similarity to the wild-type, with SAGE-Prot used to generate new proteins. The length score measures the deviation from a reference protein length, assigning a score of 1.0 for exact matches and penalizing differences proportionally. The similarity score, applied in similarity tasks, assigns a score of 1.0 if a predefined threshold is exceeded. Here, the similarity score reflects the assumption that generated proteins retain a wild-type-like structure, not a filter for low-similarity proteins. For binding affinity and thermal stability, length and similarity scores were compared using the immunoglobulin B1 binding domain of protein G (GB1) from \textit{Streptococcus group G} (\textit{GGS}, UniProt ID: P19909 and reference length: 56). For enzymatic activity and protein solubility, comparisons were based on TEM-1 $\beta$-lactamase from \textit{Escherichia coli} (\textit{E. coli}, UniProt ID: P62593 and reference length: 286). Property predictions were conducted only when the length score was 1.0. Finally, the SPO score was calculated as the sum of length, similarity, and property scores.

Multiple-property optimization (MPO) tasks aimed to simultaneously maximize two protein properties. Similar to SPO tasks, MPO considered length and similarity scores relative to the wild-type. Property scores were normalized with a maximum threshold, assigning a value of 1.0 when the score exceeded the threshold. The final score was calculated as the sum of all scores. Firstly, MPO for GB1 was designed to enhance binding affinity and thermal stability. Thresholds were set at 30 for binding affinity and 2 for thermal stability. Secondly, MPO for TEM-1 focused on improving enzymatic activity and protein solubility, using thresholds of 3.5 and 1.5, respectively. Furthermore, curriculum learning (CL) was introduced to enable effective training for length and property scores in MPO tasks. Based on previous results, 2000 samples per iteration were distributed over 50 iterations, focusing solely on fine-tuning without generation and evaluation.

\subsection{Quantitative Structure-Property Relationship (QSPR)}

The binding affinity datasets for GB1 were sourced from Olson \textit{et al}. \cite{ref28} and Wu \textit{et al}. \cite{ref29}, while the thermal stability datasets were obtained from Nisthal \textit{et al}. \cite{ref30}. The enzymatic activity datasets for TEM-1 were sourced from Firnberg \textit{et al}. \cite{ref31}, while the protein solubility datasets were obtained from Klesmith \textit{et al}. \cite{ref32}. All datasets are summarized in Table~\ref{tab:table_1}. The thermal stability is expressed as inverse thermal stability, where negative values indicate instability, and positive values represent stability. Consequently, for these four properties, larger positive values signify better performance. The maximum and minimum values for each dataset are as follows: binding affinity single ranges from 0.0021 to 5.0219 (wild-type: 1.0), binding affinity from 0.0 to 25.0, thermal stability from -4.3391 to 1.5759 (wild-type: 0.0), enzymatic activity from 0.0008 to 2.9024 (wild-type: 1.0), and protein solubility from -1.904 to 1.21 (wild-type: 0.0). A fixed random seed was utilized to perform a stratified split for 5-fold cross-validation based on the y-values into quintiles to ensure a balanced distribution.
The protein structure for \textit{GGS} GB1 was collected from the Protein Data Bank \cite{ref33} (PDB ID: 2GI9 \cite{ref34}), while that for \textit{E. coli} TEM-1 was predicted by AlphaFold (version 2) \cite{ref35}. Hydrogen atoms were added to the protein structures at pH 7.4 and their positions were optimized with the PROPKA implemented in the Schrödinger suite (ver. 2022-4) \cite{ref36}. The restrained energy minimization was performed with OPLS4 in the Schrödinger program within 0.3 Å root-mean-squared deviation \cite{ref37}. The distances between the two residues were measured with a single-linkage distance to make distance maps.
To generate numerical features for proteins, we utilized 8 protein descriptors (Onehot, PCgrades, Extended PCgrades, ESM-1b, ESM-1v, ESM-2, TAPE, and PCspairs). Onehot, PCgrades, and extended PCgrades are sequence-based descriptors. Onehot represents amino acids as a 20-dimensional vector. PCgrades compresses single amino acid properties using principal component analysis (PCA), resulting in 13 features that capture key physicochemical characteristics \cite{ref07}. Extended PCgrades is an expansion of PCgrades, including additional principal components to explain 100\% of the variance, resulting in 21 features. ESM-1b, ESM-1v, ESM-2, and TAPE are NLP-based sequence descriptors. ESM-1b has 650 million parameters and generates 1280 features using the UniRef50 database \cite{ref09}. ESM-1v shares the same architecture but is trained on the UniRef90 database, uses an ensemble of five models, and supports zero-shot inference for unseen classes \cite{ref10}. ESM-2, the successor to ESM-1b, improves architecture and training, scaling up to 15 billion parameters for better structure prediction \cite{ref11}. TAPE produces 768 features from the Pfam database \cite{ref12}. These descriptors capture evolutionary patterns in protein sequences, enabling applications in structure prediction and functional annotation. PCspairs, a structure-based amino acid pairwise descriptor, captures key information on contact potentials, water-mediated interactions, and protein-protein interactions \cite{ref07}.

To develop QSPR models for proteins, regression algorithms such as Random Forest (RF), Light Gradient Boosting Machine (LGBM), and Extreme Gradient Boosting (XGB) were utilized, leveraging decision trees to reduce overfitting and variance. These methods construct decision trees sequentially, adjusting each tree based on the errors of its predecessors to improve model performance. Hyperparameter tuning for the QSPR models was performed using grid search, with two parameters optimized for RF, three for LGBM, and four for XGB, as detailed in Table S2. Specifically, RF tuned the number of trees (n\_estimators) and the maximum tree depth (max\_depth) \cite{ref38}. For LGBM, the optimized parameters included the boosting type (boosting\_type), the number of trees (n\_estimators), and the learning rate (learning\_rate) \cite{ref39}. XGB further included the booster type (booster) along with n\_estimators, max\_depth, and learning\_rate \cite{ref40}. These optimizations ensured the effective development of QSPR models for protein analysis.

\subsection{Experimental Validation of TEM-1 Variants}

TEM-1 wild-type and six top-ranked variants were cloned into pBT7-C-His expression plasmids. Target DNA sequences were amplified by polymerase chain reaction and purified using a DNA purification kit. Cell-free protein synthesis was performed using \textit{E. coli} extracts with the ExiProgen™ system, and the expressed proteins were purified by His-tag affinity chromatography. Protein concentrations were determined using a bicinchoninic acid assay, and protein expression was confirmed by SDS-PAGE. Enzymatic activities were measured at normalized protein concentrations and reported as the mean and standard deviation from duplicate experiments. Detailed experimental procedures are described in the Supporting Information.

\section{\label{sec:results} Results}

\subsection{Scoring-Assisted Generative Exploration for Proteins (SAGE-Prot)}

\begin{figure*}[ht!]
  \centering
  \includegraphics[width=1.0\textwidth]{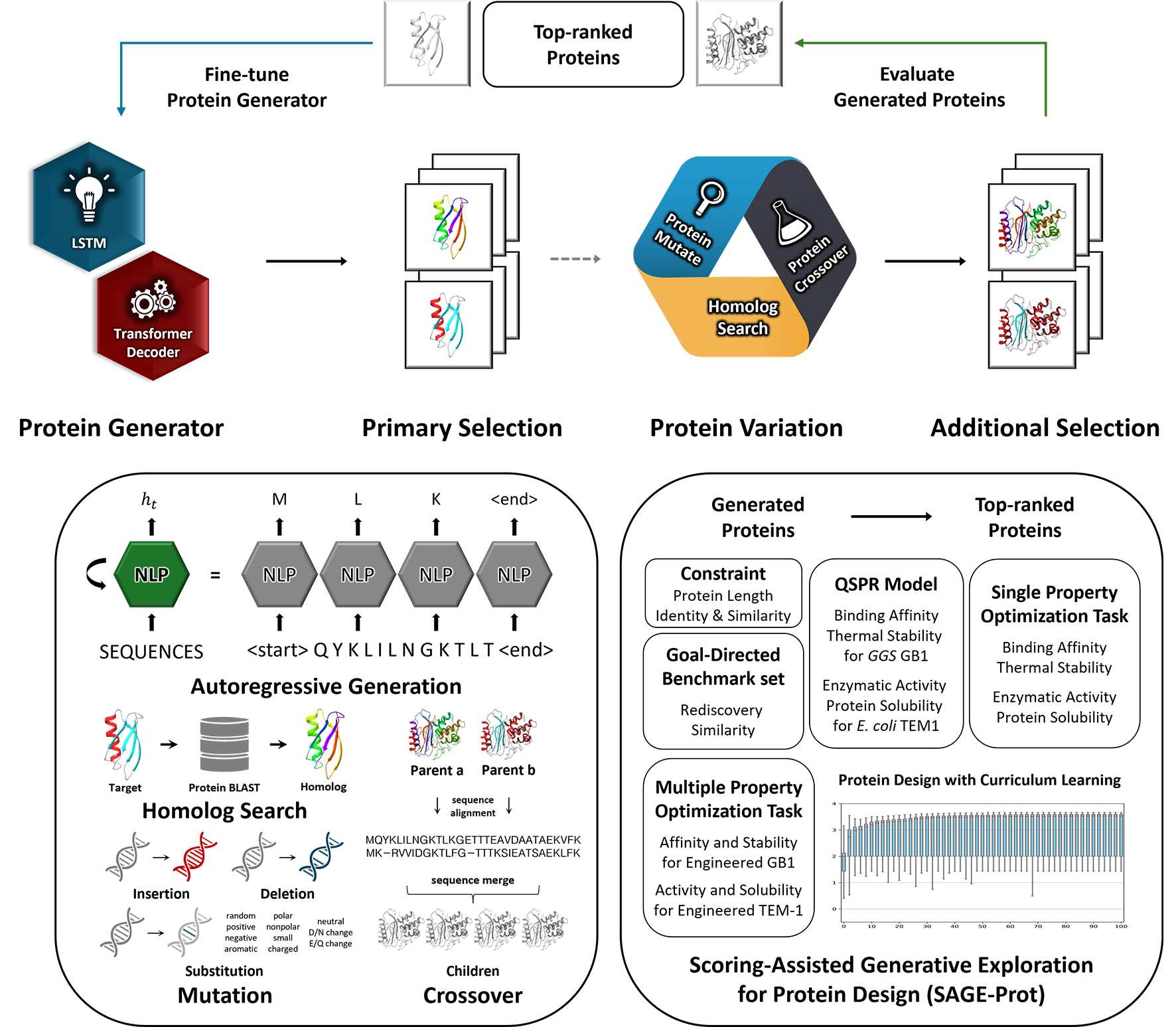}
  \caption{\label{fig:figure_1} Scoring-Assisted Generative Exploration for Protein Design (SAGE-Prot).}
\end{figure*}

SAGE-Prot is a systematic framework for optimizing protein properties through iterative fine-tuning, encompassing protein sequence generation and function evaluation. This process ultimately enables the design of novel proteins for specific purposes. As shown in Figure~\ref{fig:figure_1}, SAGE-Prot begins with protein generation, followed by evaluation and selection based on target properties, with iterative improvement achieved at each step. Initially, NLP models (LSTM and TD) are pre-trained on protein sequence datasets. These pre-trained models generate protein sequences using autoregressive modeling. Next, the generated sequences undergo variation through genetic algorithms, incorporating methods such as insertion, deletion, and substitution mutations and sequence merging through alignment, as well as retrieval-augmented generation via homolog search. In the evaluation phase, protein properties are predicted using scoring models built on various protein descriptors. Additionally, QSPR models for proteins were developed using data on binding affinity and thermal stability of GB1, as well as enzymatic activity and solubility of TEM-1. These integrations enable both single-property optimization (SPO) and multi-property optimization (MPO) within SAGE-Prot.

\subsection{Pre-training of the SAGE-Prot models with protein sequence databases}

To evaluate the suitability of SAGE-Prot for de novo protein design, its ability to generate valid proteins was tested. Protein sequences were sourced from the SwissProt and BLAST databases, resulting in three datasets: SwissProt, SwissProt-reduced, and a Custom TEM-1 dataset (Table~\ref{tab:table_1}). The NLP models in SAGE-Prot were pre-trained on these datasets using LSTM and TD algorithms. Using the pre-trained SAGE-Prot, 5000 and 10,000 proteins were generated and assessed for validity, length, uniqueness, and novelty (Table S1).

\begin{table*}[ht!]
    \centering
    \caption{Summary of Datasets Used in this work}
    \label{tab:table_1}
    \begin{ruledtabular}
    \begin{tabular}{ccccc}
        \textbf{Class} & 
        \multicolumn{2}{c}{\textbf{Task}} & 
        \textbf{Unit} & \textbf{All set} \\
        \cline{1-5}
        \multirow{3}{*}{Pre-train} & \multicolumn{2}{c}{SwissProt} & \multirow{3}{*}{Count} & 375,456 \\
        & \multicolumn{2}{c}{SwissProt-reduced} & & 375,282 \\
        & \multicolumn{2}{c}{Custom dataset for TEM-1} & & 87,856 \\
        \cline{1-5}
        \multirow{5}{*}{QSPR} & \multirow{3}{*}{GB1} & Binding Affinity Single & \multirow{2}{*}{\shortstack{Fitness ratio\\ log\textsubscript{2}(W\textsubscript{i}/W\textsubscript{wt})}} & 1046 \\
        & & Binding Affinity & & 694,720 \\
        & & Thermal Stability & ddG (kcal/mol) & 936 \\
        \cline{2-5}
        & \multirow{2}{*}{TEM-1} & Enzymatic Activity & \multirow{2}{*}{\shortstack{Enrichment ratio\\ log\textsubscript{2}(W\textsubscript{i}/W\textsubscript{wt})}} & 5199 \\
        & & Protein Solubility & & 4998 \\
    \end{tabular}
    \end{ruledtabular}
\end{table*}

When pre-trained on all datasets, the NLP algorithms generated over 98\% valid protein sequences, demonstrating that the sequences were composed entirely of canonical amino acids. The generated protein lengths closely matched the distribution observed in the training databases. For the SwissProt and SwissProt-reduced datasets, the models achieved 100\% uniqueness and novelty, successfully generating non-redundant and entirely new protein sequences. However, for the Custom dataset, the LSTM model achieved 91\% uniqueness and 82\% novelty, while the TD model showed 79\% uniqueness and 74\% novelty, indicating some redundancy and overlap with the training sequences. Overall, both LSTM and TD models effectively produced valid protein sequences with appropriate lengths, though the LSTM model outperformed the TD model in uniqueness and novelty.

\subsection{Goal-directed Benchmarks with SAGE-Prot}

\begin{table*}[ht!]
    \centering
    \caption{Results of the SAGE-Prot Models for Goal-Directed Benchmarks}
    \label{tab:table_2}
    \begin{ruledtabular}
    \begin{tabular}{cccccccc}
        \textbf{Task} & \textbf{Protein Name} & \textbf{Length} & \textbf{GA} & \textbf{LSTM} & \textbf{TD} & \textbf{LSTM/GA} & \textbf{TD/GA} \\
        \cline{1-8}
        \multirow{10}{*}{Rediscovery} & Insulin & 110 & 0.408 & 0.508 & 0.554 & 1.000 & 1.000 \\
        & Parathyroid hormone & 115 & 0.311 & 0.472 & 0.481 & 1.000 & 1.000 \\
        & Interferon gamma & 166 & 0.295 & 0.889 & 0.378 & 1.000 & 0.837 \\
        & Interferon beta & 187 & 0.289 & 0.536 & 0.377 & 1.000 & 1.000 \\
        & Interferon alpha-2 & 188 & 0.290 & 0.523 & 0.354 & 1.000 & 1.000 \\
        & Erythropoietin & 193 & 0.295 & 0.527 & 0.390 & 1.000 & 1.000 \\
        & Caplacizumab & 259 & 0.280 & 0.513 & 0.330 & 1.000 & 0.684 \\
        & Pexelizumab & 268 & 0.271 & 0.529 & 0.335 & 1.000 & 0.583 \\
        & Asparaginase & 348 & 0.280 & 0.503 & 0.308 & 0.986 & 0.491 \\
        & Thrombopoietin & 353 & 0.280 & 0.501 & 0.357 & 1.000 & 1.000 \\
        \cline{1-8}
        \multirow{10}{*}{Similarity} & Insulin & 110 & 0.434 & 0.473 & 0.542 & 1.000 & 0.920 \\
        & Parathyroid hormone & 115 & 0.417 & 0.910 & 0.756 & 1.000 & 0.940 \\
        & Interferon gamma & 166 & 0.397 & 0.540 & 0.548 & 1.000 & 0.789 \\
        & Interferon beta & 187 & 0.396 & 0.791 & 0.477 & 1.000 & 0.844 \\
        & Interferon alpha-2 & 188 & 0.393 & 0.709 & 0.481 & 1.000 & 0.864 \\
        & Erythropoietin & 193 & 0.615 & 0.444 & 0.460 & 1.000 & 0.830 \\
        & Caplacizumab & 259 & 0.373 & 1.000 & 0.413 & 0.985 & 0.573 \\
        & Pexelizumab & 268 & 0.368 & 0.732 & 0.409 & 0.985 & 0.578 \\
        & Asparaginase & 348 & 0.375 & 0.679 & 0.402 & 0.991 & 0.455 \\
        & Thrombopoietin & 353 & 0.362 & 0.507 & 0.409 & 1.000 & 0.691 \\
        \cline{1-8}
        \multicolumn{3}{c}{Total} & 7.128 & 12.287 & 8.763 & 19.946 & 16.081 \\
        \cline{1-8}
    \end{tabular}
    \end{ruledtabular}
\end{table*}

To validate the effectiveness of SAGE-Prot in de novo design, two goal-directed benchmarks (Rediscovery and Similarity) were conducted. These benchmarks evaluated the algorithms' capabilities to rediscover the same protein (Rediscovery) and generate similar proteins (Similarity). The target proteins were derived from well-established approved protein drugs. Results are presented in Table~\ref{tab:table_2}. Five algorithms (GA, LSTM, TD, LSTM/GA, and TD/GA) were compared, incorporating both NLP and GA approaches. For the NLP algorithms, model weights were pre-trained on the SwissProt-reduced dataset, excluding the similar proteins to target proteins.

In the rediscovery task, the GA-only and NLP-only models (LSTM and TD) achieved scores of 2.999, 5.502, and 3.867, respectively, with none of the three models identifying all 10 target proteins accurately. In contrast, the LSTM/GA and TD/GA models scored 9.986 and 8.596, respectively, with LSTM/GA identifying 9 target proteins and TD/GA identifying 6. These results highlight that combining NLP and GA (as NLP/GA) outperforms their usage in accurately identifying target proteins. For the similarity task, the GA-only and NLP-only models (LSTM and TD) scored 4.129, 6.785, and 4.896, respectively, while LSTM/GA and TD/GA scored 9.960 and 7.486. Again, the combined NLP/GA models demonstrated superior performance. Comparing LSTM and TD, LSTM consistently outperformed TD in both NLP-only and NLP/GA configurations. Overall, LSTM/GA achieved the highest performance on the general goal-directed benchmark with a score of 19.946, followed by TD/GA (16.081), LSTM (12.287), TD (8.763), and GA (7.128).

Additionally, using a customized dataset that includes target proteins in the training data, rediscovery and similarity tasks were performed on \textit{GGS} GB1 and \textit{E. coli} TEM-1. Results are presented in Table S2. When target proteins were absent from the training data, the LSTM, TD, LSTM/GA, and TD/GA models scored 2.724, 2.106, 3.972, and 2.817, respectively. In contrast, scores improved to 3.841, 3.321, 4.000, and 3.982 when target proteins were included. Across all scenarios, LSTM/GA, TD/GA, LSTM, and TD achieved performance scores of 7.972, 6.899, 6.566, and 5.427, respectively. Overall, these results demonstrate that LSTM is more effective than TD in generating novel proteins, making it more suitable for de novo design. Furthermore, NLP/GA significantly outperforms GA-only and NLP-only configurations. The presence of target-like proteins in the training data further enhances generation performance. Ultimately, LSTM/GA in SAGE-Prot emerged as the best-performing model, excelling in both discovering novel proteins and exploring new, similar proteins across vast protein sequence spaces.

\subsection{GB1 Design with SAGE-Prot for Binding Affinity and Thermal Stability}

To engineer GB1 proteins using SAGE-Prot, datasets for binding affinity and thermal stability were collected from the literature \cite{ref28, ref29, ref30} (Table~\ref{tab:table_1}). QSPR models for these properties were developed through a grid search to determine optimal hyperparameters using 5-fold cross-validation (Table S3). The performance metrics of the optimal models are summarized in Table S4, with the best models selected based on R\textsuperscript{2} scores from the cross-validation sets (Table S5).

Proteins with higher binding affinity can effectively interact with their targets even at low concentrations, while those with high thermal stability maintain their structure and functionality under varying temperatures, making them ideal for industrial and clinical applications. The GB1, which binds to the immunoglobulin Fc region, is particularly suited for uses like immunoprecipitation and antibody purification. In binding affinity prediction, the PCspairs/LGBM model trained on single mutations achieved the best performance, with an R\textsuperscript{2} of 0.645 and MAE of 0.252 during 5-fold cross-validation. Among the top 3 models developed using single mutations, the PCgrades/XGB model showed the best results when applied to the entire binding affinity dataset, achieving an R\textsuperscript{2} of 0.945 and MAE of 0.105. Similarly, in thermal stability prediction, the ESM-1b/LGBM model performed best, with an R\textsuperscript{2} of 0.678 and MAE of 0.583 during 5-fold cross-validation. To design GB1 with enhanced binding affinity and thermal stability, the best QSPR models were integrated into SAGE-Prot.

\begin{figure*}[ht!]
  \centering
  \includegraphics[width=1.0\textwidth]{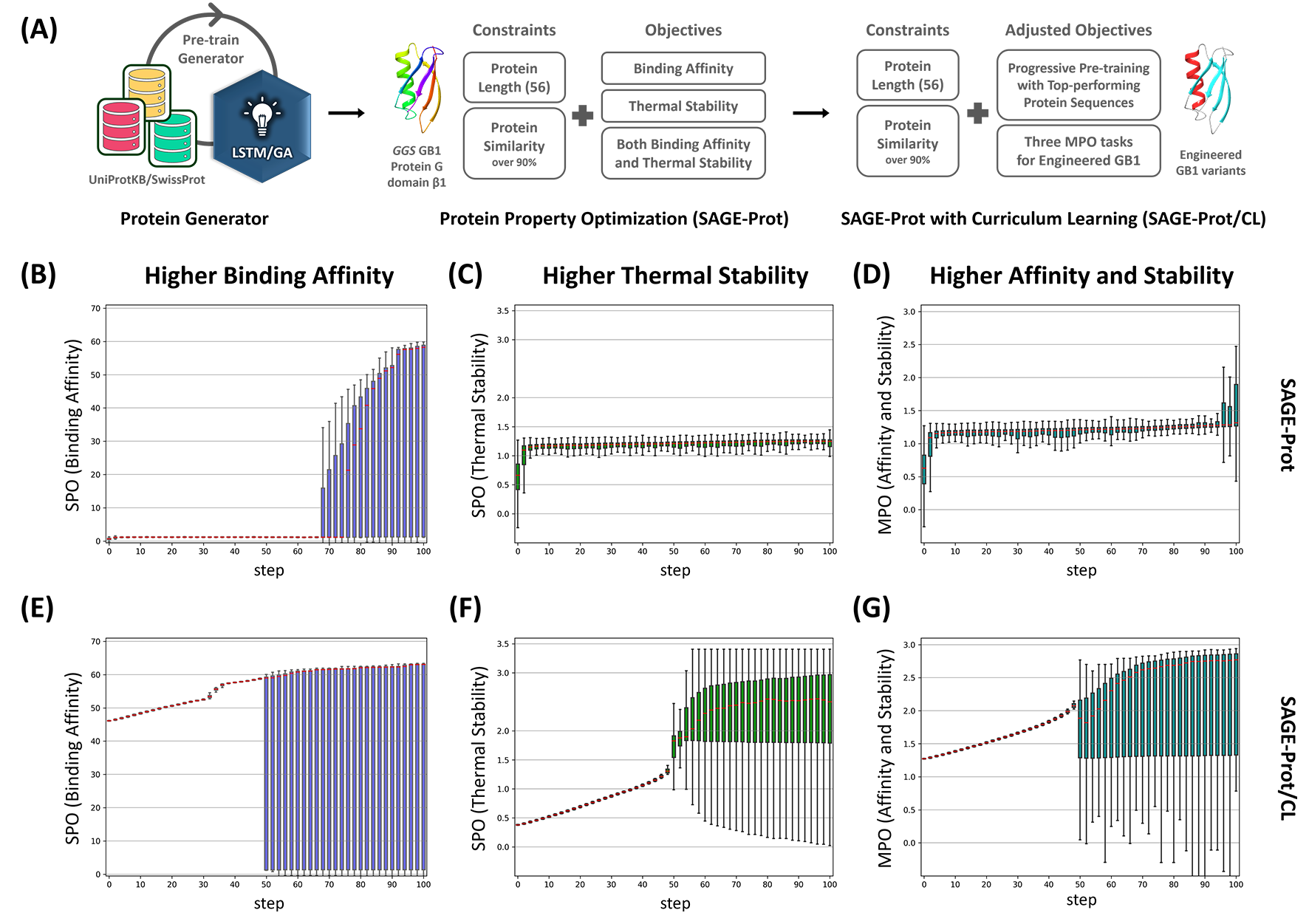}
    \caption{\label{fig:figure_2} Property Optimization of Protein GB1 Using SAGE-Prot. (A) Workflow of property optimization for Protein GB1 using SAGE-Prot. (B, C, D) Optimization trajectories of binding affinity, thermal stability, and both properties simultaneously over 100 steps using SAGE-Prot, represented as boxplots with red lines indicating the median values. (E, F, G) Optimization trajectories of binding affinity, thermal stability, and both properties simultaneously over a total of 100 steps, comprising 50 steps of curriculum learning followed by 50 steps of SAGE-Prot (SAGE-Prot/CL), also represented as boxplots with red lines indicating the median values.}
\end{figure*}

\begin{table*}[ht!]
    \centering
    \caption{Property Optimization Results of the SAGE-Prot for Protein Design}
    \label{tab:table_3}
    \begin{ruledtabular}
    \begin{tabular}{cccccc}
        \textbf{Name} & \textbf{Pre-train Dataset} & \textbf{Task} & \textbf{CL} & \textbf{SPO/MPO} & \textbf{Best SPO/MPO (properties)} \\
        \cline{1-6}
        \multirow{6}{*}{GB1} & \multirow{6}{*}{SwissProt} & \multirow{2}{*}{\shortstack{Higher\\ Binding Affinity}} & off & 59.749 & 59.865 (58.578) \\
        & & & on & 63.308 & 63.393 (62.055) \\
        \cline{3-6}
        & & \multirow{2}{*}{\shortstack{Higher\\ Thermal Stability}} & off & 1.641 & 1.777 (0.476) \\
        & & & on & 3.254 & 3.477 (1.477) \\
        \cline{3-6}
        & & \multirow{2}{*}{\shortstack{Higher\\ Both Properties}} & off & 2.423 & 2.473 (33.192 and 0.110) \\
        & & & on & 2.844 & 2.875 (41.040 and 0.405) \\
        \cline{1-6}
        \multirow{6}{*}{TEM-1} & \multirow{6}{*}{\shortstack{Custom dataset\\ for TEM-1}} & \multirow{2}{*}{\shortstack{Higher\\ Enzyme Activity}} & off & 3.570 & 3.666 (1.666) \\
        & & & on & 4.025 & 4.699 (2.699) \\
        \cline{3-6}
        & & \multirow{2}{*}{\shortstack{Higher\\ Protein Solubility}} & off & 3.071 & 3.097 (1.097) \\
        & & & on & 3.081 & 3.106 (1.106) \\
        \cline{3-6}
        & & \multirow{2}{*}{\shortstack{Higher\\ Both Properties}} & off & 2.898 & 2.995 (1.110 and 1.016) \\
        & & & on & 2.970 & 2.991 (1.169 and 0.985) \\
 
    \end{tabular}
    \end{ruledtabular}
\end{table*}

LSTM/GA in SAGE-Prot outperformed other models in goal-directed benchmarks. Pre-training on the full SwissProt database was more effective than using the SwissProt-reduced dataset for identifying proteins similar to \textit{GGS} GB1. Using SAGE-Prot, we designed GB1 optimization tasks for single-property optimization (SPO) and multiple-property optimization (MPO) to enhance binding affinity, thermal stability, or both (Figure~\ref{fig:figure_2}A). Generated proteins were evaluated based on property-specific scores, combined with a fixed protein length of 56 residues and similarity thresholds of 90\%. Iterative fine-tuning was performed over 100 generations, with the final score calculated as the average of the top 100 variants. SAGE-Prot results for the SPO and MPO tasks in the GB1 design are shown in Figures~\ref{fig:figure_2}B–~\ref{fig:figure_2}D and summarized in Table~\ref{tab:table_3}.

In the binding affinity SPO task, SAGE-Prot began generating variants with an SPO score exceeding 30.0 from step 67 onward, progressively improving the median SPO scores until step 100. By the final step, it achieved an SPO score of 59.749, with the top-performing variant reaching 59.865 and a maximum predicted affinity of 58.578 (Figure~\ref{fig:figure_2}B). Conversely, in the thermal stability SPO task, the median SPO showed an initial fluctuation at step 1 before gradually increasing, ultimately reaching 1.641 (Figure~\ref{fig:figure_2}C). The highest-scoring variant attained 1.777 at step 92, with a peak predicted stability of 0.476. For the MPO task, which optimizes both affinity and stability, SAGE-Prot followed a trajectory similar to that of the thermal stability SPO task. After an early fluctuation at step 1, the median MPO score steadily improved, culminating in a final score of 2.423 (Figure~\ref{fig:figure_2}D). The best variant achieved an MPO score of 2.473 at step 100, with a predicted affinity of 33.192 and stability of 0.110.

While SAGE-Prot improved median scores through iterative fine-tuning, it first had to undergo initial steps to satisfy constraints on protein length and similarity. To accelerate this process, we implemented curriculum learning (CL), a two-phase approach. First, we selected the top 100,000 sequences from the previously generated SAGE-Prot results and progressively fine-tuned the NLP model with the lowest-performing 2000 sequences per iteration for 50 iterations. This was followed by 50 iterations of iterative fine-tuning with generation and evaluation steps. The results of the SPO and MPO tasks using SAGE-Prot/CL are shown in Figures~\ref{fig:figure_2}E–~\ref{fig:figure_2}G. SAGE-Prot/CL generated variants with relatively higher scores from the first-generation step (step 51) compared to when CL was absent.

SAGE-Prot/CL generated variants with relatively higher scores from the first-generation step (step 51) and consistently elevated the median scores compared to the absence of CL. In the binding affinity SPO task, SAGE-Prot/CL achieved an SPO score of 63.308 (Figure~\ref{fig:figure_2}E) and identified the best variant at step 98 with a score of 63.393, which exhibited the highest predicted affinity of 62.055, an increase of 3.477 compared to the absence of CL. Next, in the thermal stability SPO task, SAGE-Prot/CL achieved an SPO score of 3.254 (Figure~\ref{fig:figure_2}F). The best variant, identified at step 52, reached 3.477, with the maximum stability improving to 1.477 (+1.001). In the MPO task, SAGE-Prot/CL achieved an MPO score of 2.844, while the best variant, identified at step 94, attained a score of 2.875 (Figure~\ref{fig:figure_2}G). This variant exhibited a predicted affinity of 41.040 (+7.848) and a stability of 0.405 (+0.295). Overall, CL effectively guided SAGE-Prot, yielding superior GB1 designs with enhanced property scores compared to training without CL.

\subsection{TEM-1 Design with SAGE-Prot for Enzymatic Activity and Protein Solubility}

\begin{figure*}[ht!]
  \centering
  \includegraphics[width=1.0\textwidth]{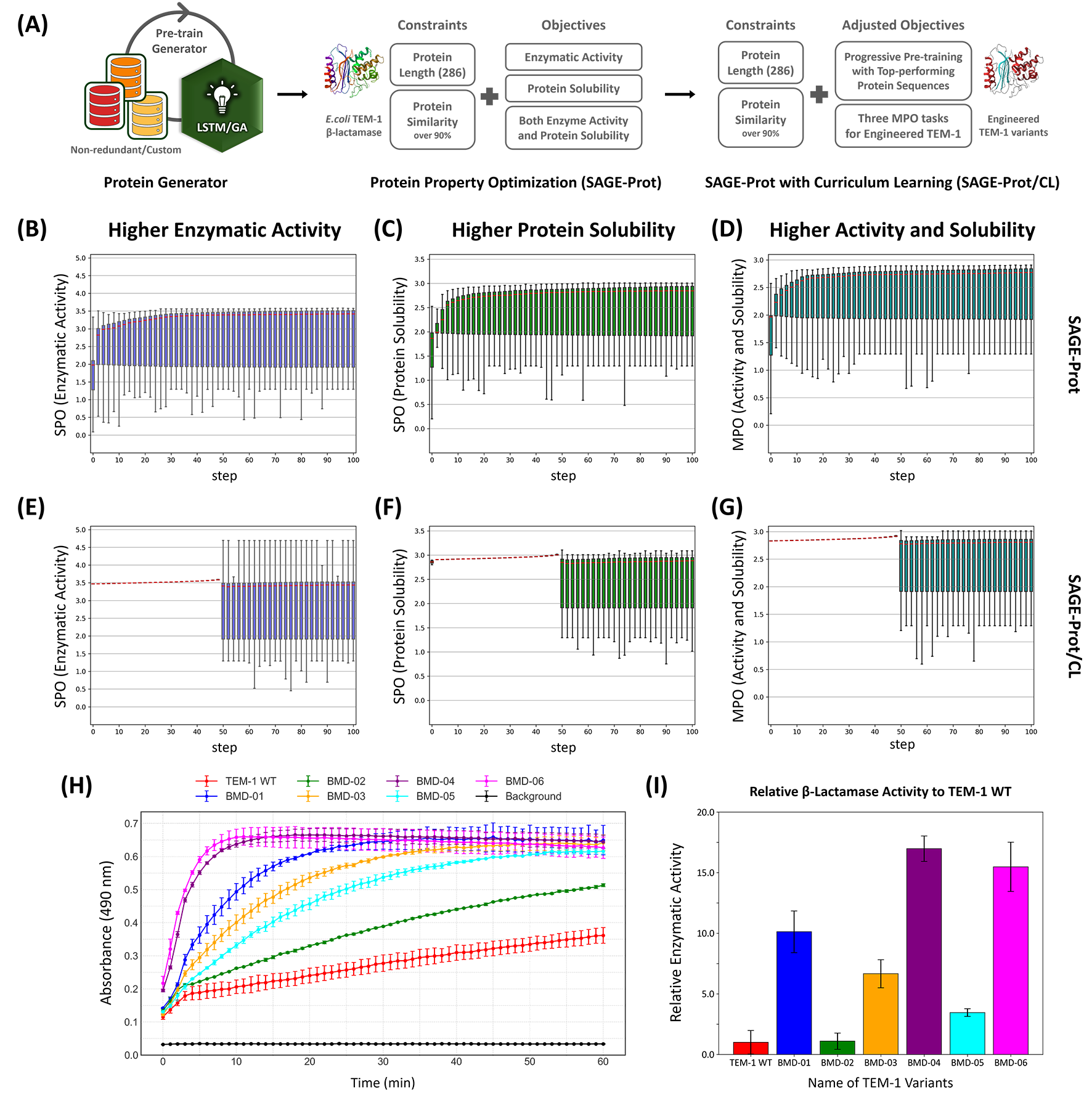}
  \caption{\label{fig:figure_3} Property Optimization of TEM-1 $\beta$-Lactamase Using SAGE-Prot. (A) Workflow of property optimization for TEM-1 $\beta$-lactamase using SAGE-Prot. (B, C, D) Optimization trajectories of enzymatic activity, protein solubility, and both properties simultaneously over 100 steps using SAGE-Prot, represented as boxplots with red lines indicating the median values. (E, F, G) Optimization trajectories of enzymatic activity, protein solubility, and both properties simultaneously over a total of 100 steps, comprising 50 steps of curriculum learning followed by 50 steps of SAGE-Prot (SAGE-Prot/CL), also represented as boxplots with red lines indicating the median values. (H) Kinetic flow curves comparing the enzymatic activities of the six top-ranked TEM-1 $\beta$-lactamase variants and the wild-type TEM-1 in \textit{E. coli}. (I) Relative $\beta$-lactamase activities of the six top-ranked variants compared to wild-type TEM-1 (normalized to 1), measured at 1 minute and 5 minutes.}
\end{figure*}

Following a similar approach to the GB1 design tasks, we applied SAGE-Prot to engineer TEM-1 with 286 residues. To achieve this, we curated datasets on enzymatic activity and protein solubility from existing literature (Table~\ref{tab:table_1}) \cite{ref31, ref32}. A grid search was used to build QSPR models, with hyperparameters optimized via 5-fold cross-validation (Table S3). The best-performing models were selected based on R\textsuperscript{2} scores from the cross-validation sets.

Enzymes with higher activity catalyze reactions more efficiently within a given timeframe, while those with greater solubility exhibit enhanced expression, making them more favorable for industrial and clinical applications. In enzymatic activity prediction, the ESM-2/LGBM model demonstrated the highest performance, achieving an R\textsuperscript{2} of 0.699 and an MAE of 0.184. Similarly, for protein solubility prediction, the same model performed best, yielding an R\textsuperscript{2} of 0.509 and an MAE of 0.273 in 5-fold cross-validation. These optimized QSPR models were integrated into SAGE-Prot to design TEM-1 variants with improved enzymatic activity and protein solubility.

Benchmarking results from SAGE-Prot with LSTM/GA showed that pre-training on a custom TEM-1 dataset was more effective than using the SwissProt-reduced database for identifying \textit{E. coli} TEM-1 and generating similar proteins. Using SAGE-Prot, we formulated SPO and MPO tasks to enhance enzymatic activity, protein solubility, or both (Figure~\ref{fig:figure_3}). Generated proteins were evaluated based on property-specific scores, with a fixed length of 286 residues and similarity thresholds of 90\%. Iterative fine-tuning was conducted over 100 iterations, and the results are summarized in Table~\ref{tab:table_3}. The outcomes of SAGE-Prot for the SPO and MPO tasks in the TEM-1 design are presented in Figures~\ref{fig:figure_3}B–~\ref{fig:figure_3}D.

In the enzymatic activity SPO task, SAGE-Prot generated variants with an SPO score of 2.0 starting from step 2. Unlike in the GB1 design, it adjusted protein length and similarity scores relatively quickly during generation. The median SPO scores in the task gradually increased to step 100, reaching 3.570 (Figure~\ref{fig:figure_3}B). The best variant, identified at step 66, achieved a score of 3.666 with the highest predicted activity of 1.666. Similarly, in the protein solubility SPO task, SAGE-Prot continuously enhanced the median SPO scores, reaching 3.071 (Figure~\ref{fig:figure_3}C). The best variant, identified at step 94, showed a score of 3.097, with the highest predicted solubility of 1.097. In the MPO task, SAGE-Prot exhibited a steady upward trend, achieving an MPO score of 2.898 (Figure~\ref{fig:figure_3}D). The best variant emerged at step 99 with a score of 2.995, alongside a predicted activity of 1.110 and solubility of 1.106.

Fine-tuning through CL in the GB1 design enhanced the property optimization process using SAGE-Prot by accelerating convergence and improving performance. Given these benefits, we applied the same approach to the TEM-1 design. In the enzymatic activity SPO task, SAGE-Prot/CL attained an SPO score of 4.025, outperforming the CL-absent condition by 0.455 (Figure~\ref{fig:figure_3}E). The best variant, identified at step 53, achieved a score of 4.699, with the highest predicted activity rising to 2.699. In the protein solubility SPO task, SAGE-Prot/CL achieved an SPO score of 3.081, reflecting a slight increase of 0.01 (Figure~\ref{fig:figure_3}F). The best variant, found at step 73, attained a score of 3.106, with predicted solubility showing a minimal rise to 0.009. Similarly, in the MPO task, SAGE-Prot/CL attained an MPO score of 2.970, reflecting a slight increase of 0.072 (Figure~\ref{fig:figure_3}G). The best variant, identified at step 55, showed a score of 2.991, with predicted activity of 1.169 and solubility of 0.985. These results suggest that the impact of CL was relatively small in the TEM-1 design using a custom dataset, compared to the SwissProt-based GB1 design. Nevertheless, CL still contributed to an improvement in SPO and MPO scores, indicating its role in refining sequence generation even under different dataset conditions.

To experimentally validate the TEM-1 design generated by SAGE-Prot, we selected the wild-type \textit{E. coli} TEM-1 along with six top-ranked variants (BMD-01 to BMD-06) identified from the MPO task. Sequence comparison between the six variants and the wild-type TEM-1 revealed that BMD-01 through BMD-06 shared 92\%, 91\%, 93\%, 92\%, 92\%, and 91\% sequence identity, respectively, across the full length of 286 amino acid residues (Figure S1). Protein structures of the six variants, when predicted using AlphaFold-3 \cite{ref41} and aligned with the wild-type TEM-1, confirmed that they maintained folding patterns similar to the wild-type. As shown in Figure~\ref{fig:figure_3}H, all designed variants displayed enhanced $\beta$-lactamase activity relative to the wild-type. Quantitatively, BMD-01 to BMD-06 exhibited approximately 10.1-, 1.1-, 6.7-, 17.0-, 3.5-, and 15.5-fold greater enzymatic activities, respectively, thereby demonstrating markedly improved catalytic efficiency (Figure~\ref{fig:figure_3}I). Collectively, these experimental validations confirmed the enhanced properties of the top-ranked TEM-1 variants, underscoring the potential of SAGE-Prot as a powerful tool for engineering proteins with customized properties for diverse real-world applications.

\section{\label{sec:dissucsion} Discussion}

Deep generative models have significantly advanced protein design by enabling the creation of novel proteins while preserving functional and structural integrity. In this study, we developed Scoring-Assisted Generative Exploration for Proteins (SAGE-Prot), a systematic framework combining autoregressive generative models with QSPR evaluations. SAGE-Prot utilizes NLP architectures to generate diverse protein sequences, enhanced by GA operators to introduce variations. Protein generation using the combined NLP/GA approach outperformed methods relying solely on NLP or GA. QSPR-based evaluations further enhanced sequence diversity while ensuring functional relevance, enabling SAGE-Prot to optimize proteins for single-property and multi-property objectives. Using SAGE-Prot, we improved the binding affinity and thermal stability of GB1 and enhanced the enzymatic activity and solubility of TEM-1. Experimental validation confirmed that the generated proteins surpassed their wild-type counterparts, demonstrating the effectiveness of SAGE-Prot in designing tailored proteins for diverse applications. By integrating advanced generative and evaluation tools, SAGE-Prot represents a major advancement in protein engineering, offering a robust platform to address complex challenges in biotechnology.

While this study demonstrates that SAGE-Prot, accelerated by CL, enables generative exploration in protein design, several limitations remain. Since SAGE-Prot operates through iterative cycles of generation and evaluation, its performance is inherently constrained by the weakness of both states. First, although NLP-GA outperformed NLP-only and GA-only in rediscovery and similarity tasks, its effectiveness declines for proteins longer than about 300 residues, failing to identify target proteins or generate 100 similar sequences. Addressing this limitation may require a larger pre-training dataset and more advanced NLP models with increased parameters. Second, in evaluating protein properties, while various descriptors were employed, prediction performance (R\textsuperscript{2}: 0.509–0.699) remained suboptimal, except for GB1 binding affinity, which benefited from abundant data. The limited accuracy of the QSPR model hinders precise extrapolation across the protein space, restricting the scope of generative exploration. Enhancing performance may necessitate expanding the dataset via double mutations or designing novel protein descriptors to extract more meaningful features from the same data. Third, despite using only top-performing sequences from previous iterations in CL, the results after 50 iterations surpassed those at 100 iterations. This suggests that refining CL by segmenting training stages or redefining scoring criteria could further enhance performance. Additionally, while landmark databases were utilized for homolog searches due to computational constraints, incorporating non-redundant or curated datasets could extend SAGE-Prot to retrieval-augmented generation. Such advancements will elevate the capabilities of generative modeling in protein design, ultimately providing a robust platform for engineering novel proteins to address complex industrial challenges.

This study demonstrates the power of generative models in protein design and engineering, with SAGE-Prot effectively optimizing multiple properties through iterative exploration and evaluation. By integrating autoregressive generation, genetic algorithms, and QSPR-based assessments, SAGE-Prot enables precise navigation of the sequence-function landscape, surpassing traditional design approaches. Experimental validation confirms its ability to generate proteins with enhanced functionality, highlighting its potential as a transformative tool for biotechnology and medicine. Despite existing limitations, such as sequence length constraints and QSPR model accuracy, future advancements in pretraining datasets, descriptor engineering, and retrieval-augmented generation would further enhance its performance. As deep generative approaches continue to evolve, SAGE-Prot represents a significant step toward computational protein design, paving the way for innovative solutions in therapeutics, biocatalysis, and beyond.

\section*{\label{sec:author contributions}Author Contributions}
H.L. conceptualized the study, developed computational methods, supported biochemical experiments, and wrote the manuscript. G.L. conducted biochemical experiments. K.N. advised this work.

\section*{\label{sec:declarations}Declarations}

\subsection*{Competing interests}
The authors declare no competing interest.

\subsection*{Code Availability}
All results in this work can be found at \url{https://github.com/hclim0213/SAGE-Prot}.

\begin{acknowledgments}
The research was supported by the Ministry of Trade, Industry, and Energy (MOTIE), the Republic of Korea, under the project “Industrial Technology Infrastructure Program” (Project No. RS-2024-00466693).
\end{acknowledgments}



\end{document}